%% file: main.tex
\documentclass{iutbscthesis}
\usepackage{kantlipsum} 
\usepackage[table]{xcolor}
\title{Analysis on LLMs Performance for Code Summarization}

\addauthor{Salman Ahsan}{190042114}
\addauthor{Md. Muktadir Mazumder}{190042136}
\addauthor{Md. Ahnaf Akib}{190042137}

\supervisor{Lutfun Nahar Lota}{Assistant Professor}{Department of Computer Science and Engineering}

\addcosupervisor{Md Farhan Ishmam}{Lecturer}{Department of Computer Science and Engineering}

\department{Department of Computer Science and Engineering}

\program{BACHELOR OF SCIENCE IN SOFTWARE ENGINEERING}

\defensedate{5}{July}{2024}

\begin{document}


\coverpage

\pagenumbering{roman}

\titlepage

\declarationofcandidate


\tableofcontents
\listoffigures
\listoftables

\clearpage

\begin{abbreviations}
    \abbr{NLP}{Natural Language Processing}
    \abbr{LLM}{Large Language Model}
    \abbr{PIP}{PIP Installs Packages}
    \abbr{LoRa}{Log-odds Ratio Attention}
    \abbr{LSTM}{Long Short-Term Memory}
    \abbr{GNN}{Graph Neural Network}
    \abbr{AST}{Abstract Syntax Tree}
    \abbr{RNN}{Recurrent Neural Network}
    \abbr{GN}{Graph Network}
    \abbr{SCG}{Syntax Code Graph}
    \abbr{BLEU}{Bilingual Evaluation Understudy}
    \abbr{METEOR}{Metric for Evaluation of Translation with Explicit Ordering}
    \abbr{ROUGE}{Recall-Oriented Understudy for Gisting Evaluation}
    \abbr{DG}{Dynamic Graph}
    \abbr{BPE}{Byte Pair Encoding}
    \abbr{SFT}{Supervised Fine-Tuning}
    \abbr{SiT}{Structure Induced Transformer}
    
\end{abbreviations}

\begin{acknowledgement}





We would like to express our gratitude towards IUT authority for providing assistance required to implement our proposed system. We are indebted to our supervisor, Lutfun Nahar Lota for providing us with insightful knowledge and guiding us at every stage of our journey. 
Also, we would like to thank our co-supervisor Md Farhan Ishmam for his special contributions.

Finally, we would like to express our heartiest appreciation towards our family members for their continuous support, motivation, suggestions, and help, without which we could not have achieved the scale of implementation that we have achieved.
\end{acknowledgement}

\input{chapters/abstract}
\pagenumbering{arabic}

\input{chapters/introduction}
\input{chapters/relatedworks}

\input{chapters/methodology}

\input{chapters/result}
\input{chapters/discussion}
\printbib




\end{document}

%% file: chapters/abstract.tex
\begin{abstract}
The goal of code summarizing is to produce concise source code descriptions in natural language. Deep learning has been used more and more recently in software engineering, particularly for tasks like code creation and summarization. Specifically, it appears that the most current Large Language Models with coding perform well on these tasks. 


Code summarization has evolved tremendously with the advent of Large Language Models (LLMs), providing sophisticated methods for generating concise and accurate summaries of source code. Our study aims to perform a comparative analysis of several open-source LLMs, namely LLaMA-3, Phi-3, Mistral, and Gemma. These models' performance is assessed using important metrics such as BLEU\textsuperscript{\ref{BLEU}} and ROUGE\textsuperscript{\ref{ROUGE-L}}.

Through this analysis, we seek to identify the strengths and weaknesses of each model, offering insights into their applicability and effectiveness in code summarization tasks. Our findings contribute to the ongoing development and refinement of LLMs, supporting their integration into tools that enhance software development and maintenance processes.


\vspace{\baselineskip}
\end{abstract}

%% file: chapters/introduction.tex
\chapter{Introduction}\label{chapter:introduction}
The application of Natural Language Processing (NLP) techniques for automated program understanding, production, and retrieval is becoming more popular, as these tasks have the potential to enhance code accessibility. One common activity is Code Summarization, which is essentially translating code into Natural Language \cite{barone2017parallel}. This work is crucial because software developers can become much more productive if they can automatically generate code summaries or doc strings.

Although existing models perform impressively well in code summarization, it is crucial to assess how much the structure and semantics of the code are understood by these models. To make their code more legible by humans, software writers frequently use English terms as the names of variables, functions, and data structures.

This study compares a number of open-source Large Language Models (LLMs)—namely, LLaMA-3, Phi-3, Mistral, and Gemma—that are utilized for code summarization in order to ease these worries. Key performance indicators like BLEU, F1 Score, Precision, Accuracy, and ROUGE are used to assess these models' performance. By means of this study, we want to evaluate the benefits and drawbacks of every model, providing valuable perspectives on their suitability and efficiency in code summarizing assignments. Our research supports the further improvement and development of LLMs and their incorporation into tools that improve the processes involved in software development and maintenance.

\section{Overview}




 
\subsection*{What's Source Code Summarization ?}
A brief description of the code's function in simple language may be found in the source code summary \cite{tang2022ast}. Writing concise explanations of code in normal language is known as source code summarization \cite{leclair2020improved}. Code summarization, sometimes referred to as code commenting, is a textual explanation of the role that specific identifiers play in computer systems. Stated differently, code summary aids in program comprehension by providing a natural language explanation of the logic and functionality of the source code \cite{zhang2022survey}. These descriptions, which offer context and functional insights, assist in improving the code's comprehensibility and accessibility for developers and other stakeholders.

In order to visually represent the topic source code summarization; some examples are provided below:

\subsection*{Use Case 01}
\vspace{0.3cm}
\begin{minipage}{0.45\textwidth}
 In Figure~\ref{fig:s1}, we can see a simple block of code in java programming language, specifically a function named \textit{addNumbers}, which takes two integer numbers as parameters and returns the result of their addition. So, the summarized text of the code will be simply "\textit{returns the sum of
two integers}"~in natural language which is readable to human.

\end{minipage}
 \hfill
\begin{minipage}{0.5\textwidth}
\begin{figure}[H]
 \includegraphics[width=1\linewidth]{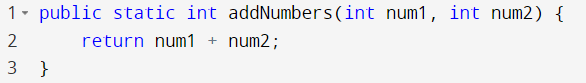}
        \caption{A Simple block of code or function}
        \label{fig:s1}
\end{figure}

\end{minipage}

\subsection*{Use Case 02}
\vspace{0.3cm}

\begin{minipage}{0.45\textwidth}

 In Figure~\ref{fig:s2}, if we follow the red, green and blue lines on the simple block of java code, we observe their combination to form specific keywords. These extracted keywords play a pivotal role in generating the code summary. Consequently, the resulting summarized text in natural language is "\textit{contains ignore case}," which remains comprehensible to humans.

\end{minipage}
 \hfill
\begin{minipage}{0.5\textwidth}
\begin{figure}[H]
 \includegraphics[width=1\linewidth]{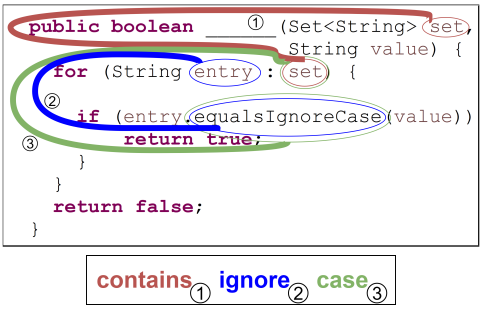}
        \caption{Combining keywords to generate code summaries}
        \label{fig:s2}
\end{figure}

\end{minipage}

\subsection*{The importance of source code summarization}
Summarizing source code is essential because it makes code more readable and manageable by giving clear, succinct explanations of its functionality. This helps developers navigate and grasp vast codebases more rapidly, increasing productivity and decreasing the amount of time spent trying to understand complex code structures.

Source code summarization has several real-world applications:

\begin{enumerate}
    \item \textbf{Code Documentation:} Automatically generated summaries can improve the documentation of software projects, making it easier for developers to understand and contribute to the codebase.
    \item \textbf{Code Review:} Summaries can assist code reviewers by providing quick insights into the functionality of code snippets, streamlining the review process.
    \item \textbf{Bug Tracking and Fixing:} Summaries help developers quickly grasp the purpose of code sections when diagnosing and fixing bugs, enhancing efficiency.
    \item \textbf{Learning and Onboarding:} New team members can benefit from summarized code to quickly get up to speed with the project's structure and functionality, accelerating the onboarding process.
    \item \textbf{Search and Retrieval:} Summarized code can enhance code search engines, making it easier to find relevant code snippets based on natural language queries.
    \item \textbf{Software Maintenance:} Summarized code assists in maintaining large codebases by providing clear explanations of code functionality, making updates and modifications more manageable.
    \item \textbf{Code Comprehension:} Summaries improve overall code comprehension, enabling developers to understand complex code structures more efficiently and effectively.
\end{enumerate}

\section{Motivation and Scope}
Efficient tools for understanding and managing code are crucial due to the increasing complexity of software systems. Source code summarization enhances productivity by providing concise, human-readable summaries, improving documentation, aiding in code reviews, bug fixing, and onboarding new developers. This research can lead to better software maintenance, faster development cycles, and reduced project costs.

Researchers are applying encoder-decoder architectures to software engineering as a result of their introduction in natural language processing \cite{sutskever2014sequence} (both Transformer-based \cite{cho2014learning} and recurrent \cite{hendrycks2021measuring}). Making code summaries is one significant application \cite{shido2019automatic}. For example, a code summary tool can be used to write documentation or comprehend legacy code. With the arrival of LLMs, or large language models, working programmers now have a lot more chances to use deep learning-based technologies. Both closed models (like GPT-4\cite{achiam257532815gpt} or Gemini\cite{team2023gemini}) and open models (like CodeLlama\cite{roziere2023code}) show remarkable ability in creating natural language summaries of code and source code based on task descriptions. This technical report's primary goal is to examine how open-sourced LLMs handle source code in respect to natural language text. Previous studies have explored various techniques for source code summarization. Graph Neural Nets, Graph Attention Neural Nets, Abstract Syntax Trees, and sequence-to-sequence models have all been created, along with semantic embeddings for code snippet representation and code comment generation. Inspired by transformer based models like BERT and GPT, LLMs such as CodeBERT, GraphCodeBERT, PLBART, and CodeT5 have achieved significant advancements.

This study performs a comparative analysis of open-source LLMs, including LLaMA-3, Phi-3, Mistral, and Gemma, for code summarization. By evaluating these models on metrics like BLEU, F1 Score, Precision, Accuracy, and ROUGE-L, the research aims to identify models that generate syntactically and semantically accurate summaries, contributing to more effective software development tools. The main objective is to investigate how well open-sourced LLMs handle source code in relation to natural language text.

\section{Problem Statement}
In this study, we aim to review how open-source large language models (LLMs) perform in code explanation or summarization. This research addresses the challenge of evaluating the effectiveness of these models in generating accurate and meaningful summaries of source code. The specific objectives are to compare the performance of open-source LLMs (LLaMA-3, Phi-3, Mistral, and Gemma) on code summarization tasks and to evaluate these models using performance metrics such as BLEU and ROUGE. By identifying the strengths and weaknesses of each model in handling the semantic relationship between source code and natural language, this study aims to provide insights into which models offer the best balance of syntactic and semantic accuracy for code summarization tasks.

Furthermore, the project seeks to fine-tune these LLMs to enhance their ability to generate accurate, concise, and contextually relevant summaries for code snippets. Despite advancements in natural language processing, existing LLMs often struggle with providing precise and contextually appropriate summaries for complex code structures. This limitation hampers the efficiency of developers who rely on these models for understanding and documenting code. Therefore, by refining the LLMs, we aim to improve their comprehension of programming languages and enable them to produce summaries that are both succinct and highly relevant to the given code context.

\section{Research Challenges}
Code summarization, which involves generating concise descriptions of source code segments, faces several significant research challenges. One of the primary difficulties is the semantic gap between the low-level operations described in the code and the high-level tasks they accomplish. Bridging this gap requires a deep understanding not just of the code's syntax, but also of its semantics and the programmer's intent. Additionally, domain-specific knowledge is often necessary to accurately interpret and summarize code, particularly in specialized fields. Another challenge is the variability in programming styles and languages, which can affect the effectiveness of summarization tools. Moreover, the limited availability of high-quality, annotated datasets for training machine learning models in this area hampers progress. Lastly, ensuring the generated summaries are both accurate and useful for developers—capturing not just what the code does, but also why it does it—is a critical yet challenging aspect of research in code summarization.

\section{Contribution}
Our study adds significantly to the field of code summarization in a number of ways. First of all, it offers a thorough comparison of four Large Language Models (LLMs) that are available as open-source software, namely Gemma, Mistral, LLaMA-3, and Phi-3, with a focus on code summarization. Through a methodical assessment of these models with measures including ROUGE, F1, Precision, Accuracy, and BLEU, the study provides insightful information about the relative advantages and disadvantages of each model. 

Secondly, the research establishes a robust evaluation framework for assessing LLMs in code summarization tasks, which can serve as a benchmark for future studies, enabling consistent and objective comparisons across different models and datasets. Additionally, by analyzing the performance of various LLMs, the study identifies best practices for applying these models to code summarization, including recommendations on model selection based on specific use cases, programming languages, and code structures. 

The results also show the limitations and practical advantages of employing LLMs for code summarization in real-world software development, including increased productivity, code comprehension, and maintenance. By giving a thorough grasp of how various LLMs perform in code summarization, directing future research and useful applications, and facilitating software developers' ability to utilize LLMs' power in their work, their contributions enhance the area.

\section{Organization}
The structure of this research study is as follows: We present a thorough literature review that summarizes previous research on the function and effectiveness of LLMs in source code summarization in chapter \ref{chapter:literature}. The shortcomings and deficiencies in the state of the research are highlighted in this section. The research methodology is described in Chapter \ref{chapter:methodology}, wherein the use of Large Language Models (LLMs) - Gemma, Mistral, LLaMA-3, and Phi-3 - is discussed. Our experimental setup and dataset are also presented, and the results and analysis are presented in Chapter \ref{chapter:result}. 
Finally, in the case study and chapter \ref{chapter:discussion} address the implications of our findings, highlight their limits, and offer suggestions for further research in this area.

%% file: chapters/relatedworks.tex
\chapter{Related Works}\label{chapter:literature}
Before diving into our main reserach study on LLMs for code summarization let us first disscuss about some literatures where  Traditional Models For Code Summarization are mentioned.

\section{RNN Based Models}

\citeauthor{alon2018code2seq} \cite{alon2018code2seq} proposed a study which describes how the Sequence-to-Sequence (seq2seq) models, derived from neural machine translation (NMT), handle source code as a sequence of tokens and attain state-of-the-art performance on these tasks. The paper presents Code2Seq, a novel approach that enhances source code encoding by making use of the syntactic features of programming languages. Their methodology represents a code snippet as the set of compositional pathways in its abstract syntax tree (AST), using attention to select the pertinent paths during decoding. The authors have demonstrated the effectiveness of their method on two tasks, two programming languages, and four datasets including up to 16 million cases. Compared to previous models designed specifically for programming languages and current state-of-the-art NMT models, the model given in this research performs significantly better. However, the approach has certain limitations. Firstly, RNNs, the backbone of the model, are known to be slow to train, often requiring truncated versions of backpropagation and demanding significant computational resources.

\begin{figure}[h]
    \centering
    \includegraphics[width=0.6\textwidth]{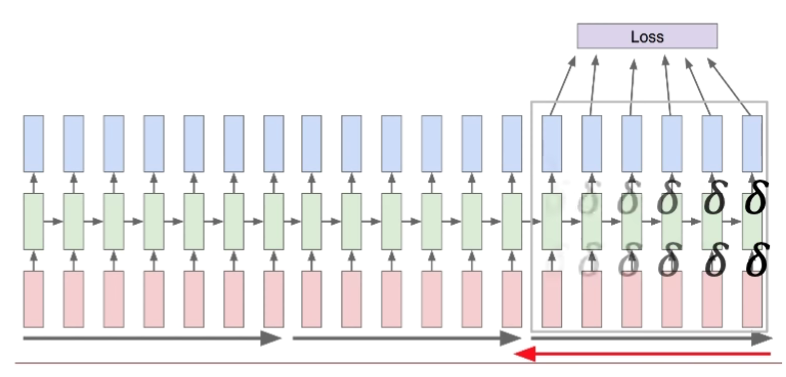}
    \caption{Illustration of Vanishing and Exploding Gradients in RNNs}
    \label{fig:rnn} 
\end{figure}

Moreover, these models are prone to vanishing and exploding gradients, especially when dealing with long sequences, as illustrated in Fig~\ref{fig:rnn}. Additionally, the architecture may struggle with large codebases, given its sequential processing nature and the challenges associated with capturing long-range dependencies effectively. In summary, while Code2Seq represents a significant advancement in source code processing, particularly in the context of seq2seq models, its applicability may be hindered by training complexities, gradient instability, and scalability issues when handling extensive codebases.

Furhtermore, \citeauthor{iyer2016summarizing} \cite{iyer2016summarizing} were the first to propose CODE-NN. Subsequent studies added more structural and syntactic information to deep learning models to increase the accuracy of code modeling. The structure-based traversal (SBT) approach was proposed by \citeauthor{hu2018deep} and \cite{hu2018deep}. Its function is to traverse the AST and convert its nodes into sequences that are appropriate for an RNN encoder. An API Sequences Encoder was added to their model by another study \cite{hu2018summarizing}, which maintained that code APIs provide crucial information about the functioning of the source code.

\section{Tree/GNN Based Models}

\citeauthor{leclair2020improved} \cite{leclair2020improved} presented a new approach in contrast to RNNs that uses a graph neural network (GNN) to summarize source code. The key idea is to make summarization easier by using the structure of the abstract syntax tree (AST) and the source code sequence. The authors test their strategy on a dataset of $2.1$ million Java method-comment pairings. The model architecture is covered in full in the paper \cite{leclair2020improved}, which also explains how the GNN and recurrent neural networks (RNNs) are integrated to encode the source code and AST. The outcomes demonstrate a significant improvement in code summarization quality, which is attributable to the GNN's effective representation of the code's structural information. Their approach is based on the graph2seq model, with certain adjustments made to better fit the concept into a software engineering setting. In short, they coupled the GNN-based encoder of graph2seq to model the AST of each subroutine with the RNN-based encoder used by \citeauthor{leclair2019neural} \cite{leclair2019neural} to model the subroutine as a sequence.

The main innovation of this approach is using GNNs to model the Abstract Syntax Trees (ASTs) of program subroutines. These ASTs are then paired with an RNN-based encoder to represent the subroutine as a sequence. By combining the two, code can be represented in a more organized way that nevertheless preserves its syntactic and sequential elements. The token sequence from the source code is embedded by the model, which also integrates the tokens from the AST nodes into the system. When encoding, the embedded source code token sequence is encoded using a recurrent layer, and the edges and nodes of the AST are encoded using convolutional graph neural networks (ConvGNNs). The encoder's outputs are decoded after the attention mechanism has identified the crucial tokens in the AST and source code. It forecasts the token that will follow in the summary sequence using the context vector that the decoder has produced.

The general context of the codebase could be challenging for the model to represent. While it can effectively analyze individual subroutines or short code segments, it could miss more complex patterns, dependencies, or features that are dispersed across the code. The method necessitates extensive preparation, especially when creating and managing the Abstract Syntax Tree (AST) of the code. This complexity can limit the model's scalability or efficiency, which raises the computational expense. This is especially true for environments with limited computational resources or for very large codebases.

\begin{figure}[h]
    \centering
    \includegraphics[width=0.6\textwidth]{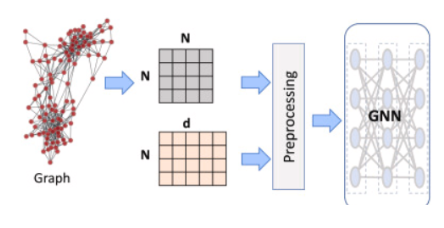}
    \caption{Preprocessing in GNN}
    \label{fig:gnn} 
\end{figure}

The challenge of precisely extracting and encoding the data from the AST into a matrix form while preserving the syntactic and semantic links seen in the source code is most likely the cause of the preprocessing step's difficulty.

A multi-way Tree-LSTM \cite{shido2019automatic} was suggested to use the ASTs' tree structures to directly model code structures. Numerous research have included code-related graphs and GNNs to improve performance for more in-depth exploitation of intra-code linkages. \citeauthor{leclair2020improved} \cite{leclair2020improved} used convolutional graph neural networks to obtain code representation directly from ASTs, whereas \citeauthor{fernandes2018structured} \cite{fernandes2018structured} constructed a graph from source code and extracted node features. In order to assist models capture more global interactions among nodes, a recent study \cite{liu2020retrieval} proposed a hybrid GNN that incorporates information from static and dynamic graphs via hybrid message passing.

\section{Transformer Based Models}

In 2017, then breakthrough obtained in tansformer based approaches shown by \citeauthor{vaswani2017attention} \cite{vaswani2017attention}  is able to determine the dependencies and relationships between various parts of the source code from the data itself. The Transformer architecture makes data processing possible in parallel, improving the model's scalability to larger datasets. By using attention mechanisms, the Transformer can better understand the context within which certain code elements appear. The model provides interpretability, possibly through mechanisms that allow visualization of attention weights to understand how the model is making its predictions.

\begin{figure}[htbp]
    \centering
    \includegraphics[width=0.4\textwidth]{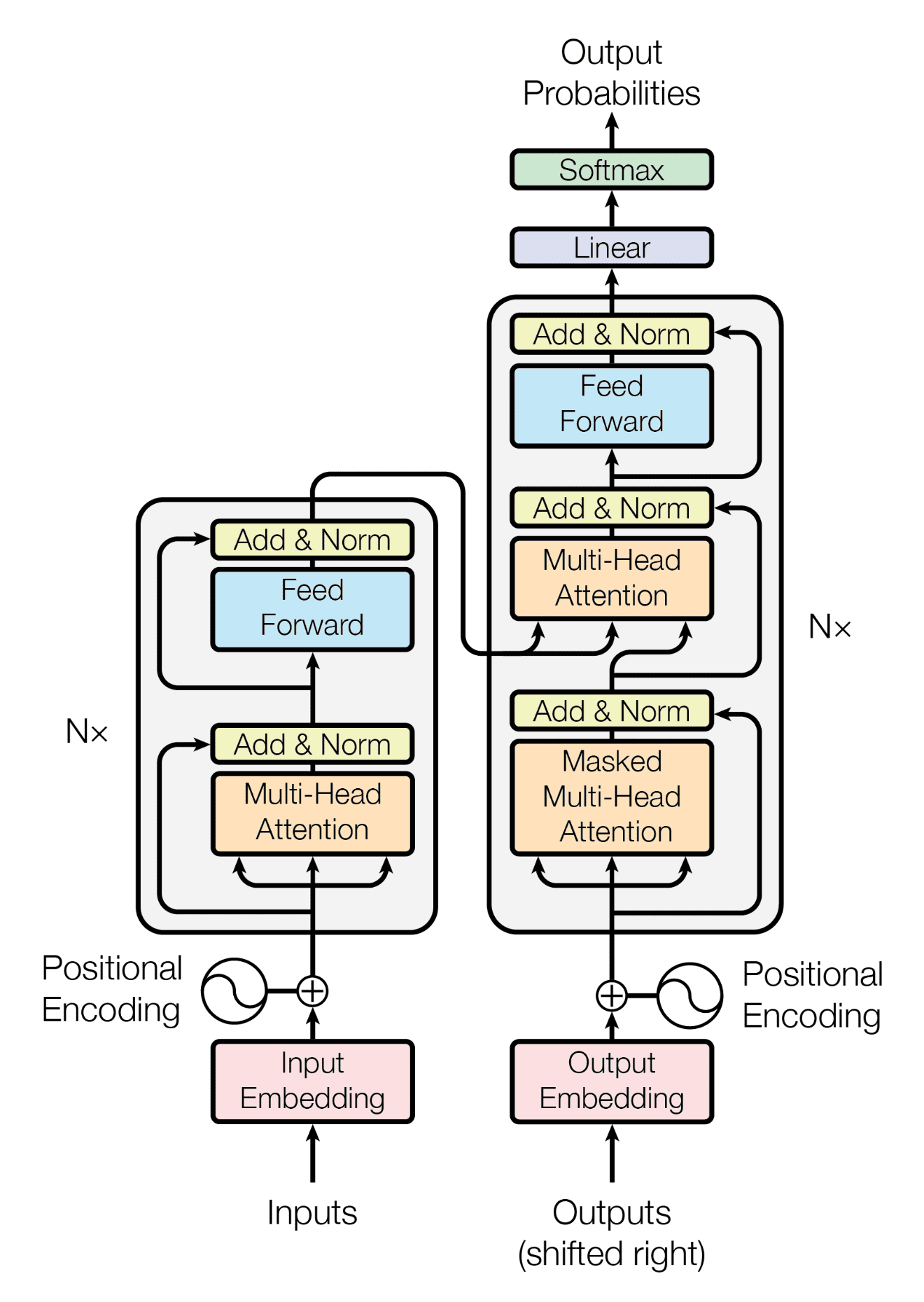}
    \caption{Transformer-based Architecture}
    \label{fig:gnn} 
\end{figure}

The model may unnecessarily attend to the same pieces of information more than once due to redundancy in the attention mechanisms. Despite the Transformer's many advantages, it is still difficult to incorporate code structure information into the Transformer model in an efficient manner.

\citeauthor{gao2021code} \cite{gao2021code} proposed a new method to integrate code structural properties into Transformer (state-of-the-art model), which they called SG-Trans. Basically, they introduce inductive bias into Transformer's self-attention module by injecting both global syntactic structure—such as the data flow graph—and local symbolic information—such as code tokens and statements. The local data and global structure are intended to disperse among the attention heads of lower layers and high layers of Transformer in order to better capture the hierarchical qualities of code. The two benchmark datasets used in the tests by the authors contained Python and Java, respectively. In particular, 87, 136 $\langle$Java method, comment$\rangle$ pairs gathered from 9, 714 GitHub repositories make up the Java dataset, while 92, 545 functions and accompanying documentation make up the Python dataset.


\citeauthor{tang2022ast} \cite{tang2022ast} introduces a more recent method by using an encoder-decoder architecture based on transformers. The source code's Abstract Syntax Tree (AST), which is extensively used for encoding structural information, is highly structured and adheres to rigid grammars. But ASTs are substantially longer than the matching source code. Current methods just feed the whole linearized AST into the encoders, ignoring the size constraint. The paper's authors contended that extracting the genuinely useful dependency relations from the lengthy input sequence is challenging due to the simplistic nature of the process. It also has a large computational overhead because every node in the AST has to give itself attention to every other node. In this paper, the authors propose AST-Trans, which uses sibling and ancestor-descendant relationships—two types of node relationships found in the AST—to encode the AST more effectively and efficiently. Based on these two relationships, it uses the tree-structured attention to dynamically assign weights for relevant nodes and exclude irrelevant nodes. The authors also suggest a quick and effective implementation to enable parallel computing for tree-structure attention. Tests are carried out on the two publicly available code summarization benchmarks, one in Python\cite{wan2018improving} and the other in Java \cite{hu2018summarizing}. The outcomes of the tests demonstrate that AST-Trans is times more efficient than standard transformers and performs noticeably better than the state-of-the-arts.

\begin{figure}[htbp]
    \centering
    \begin{subfigure}{0.49\textwidth}
        \centering
        \includegraphics[width=0.83\linewidth]{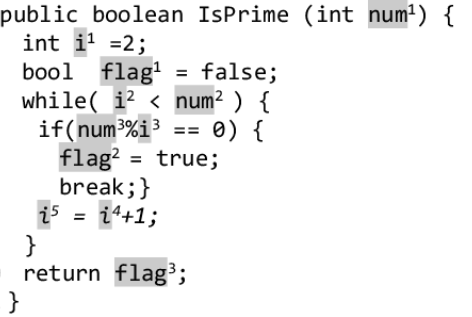}
        \caption{Code Snippet}
        \label{fig:codeSnippet}
    \end{subfigure}
    \hfill
    \begin{subfigure}{0.49\textwidth}
        \centering
        \includegraphics[width=0.99\linewidth]{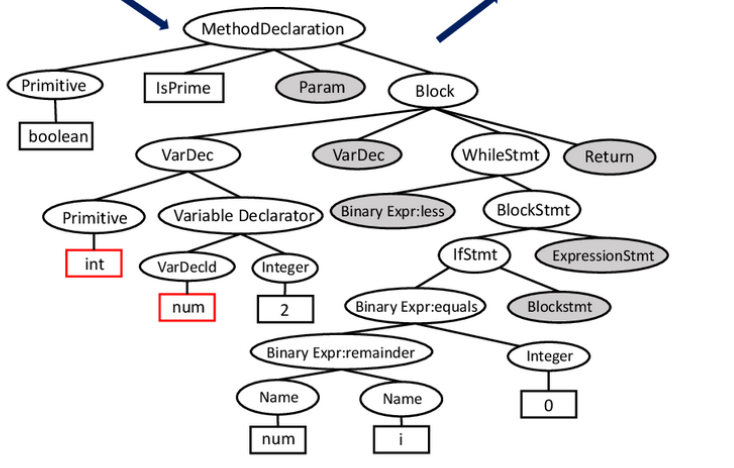}
        \caption{Abstract Syntax Tree Representation}
        \label{fig:astTree}
    \end{subfigure}   
    \caption{Representation of code snippet and its Abstract Syntax Tree}
    \label{fig:ast}
\end{figure}

The model presents an effective method for encoding source code's Abstract Syntax Tree (AST), a hierarchical depiction of the code's structure. The java code snippet's in Figure~\ref{fig:codeSnippet} and its corresponding AST in Figure~\ref{fig:astTree}, showing how the method, parameters, and body are represented in a tree structure.

The method is computationally efficient because it preserves linear complexity in spite of the difficulties involved in managing tree structures. This model's ability to handle code's tree-like structures is essential for understanding how programming languages are logically and syntactically organized.

One drawback of the approach is that, in comparison to other models, it performs worse on Python datasets.

\begin{figure}[h]
    \centering
    \includegraphics[width=0.7\textwidth]{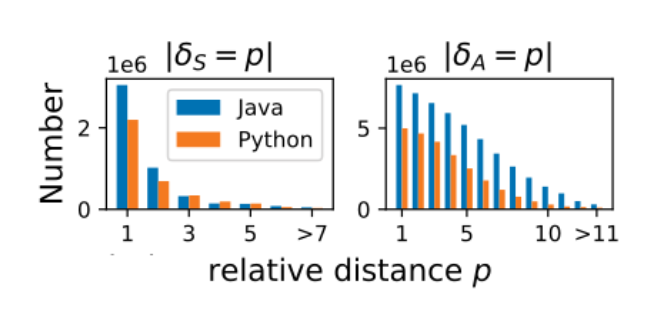}
    \caption{Performance of AST Transformer in Python and Java dataset}
    \label{fig:pAST} 
\end{figure}

This could mean that, in comparison to how it handles other languages like Java, the method performs less well when dealing with the syntax or coding patterns of Python. The histogram in Figure~\ref{fig:pAST} compares the performance of models on Java and Python datasets, showing that the current model performs less effectively on Python, as indicated by the number of occurrences of certain relative distances $\rho$.

\citeauthor{zhang2020retrieval} \cite{zhang2020retrieval} presents a novel approach The paper introduces two advanced methods for source code summarization. The first method, Rencos, combines neural machine translation (NMT) with information retrieval (IR) to enhance summary accuracy by integrating retrieved similar code snippets. The second method, Structure-induced Transformer (SiT), incorporates structural information from Abstract Syntax Trees (ASTs) into a Transformer model to capture long-term dependencies and structural relationships in the code. Both methods demonstrate significant performance improvements over existing models in standard benchmarks, although they face challenges related to training set quality and computational efficiency.

\citeauthor{wan2018improving} \cite{wan2018improving} proposes a deep reinforcement learning-based model for automatic source code summarization, introducing a hybrid representation combining lexical tokens and abstract syntax trees (ASTs). The model uses an actor-critic architecture to improve the generation and evaluation of code comments, demonstrating superior performance over existing methods. However, its effectiveness relies heavily on the quality and size of the training data, and the approach can be computationally intensive. Despite improvements, the generated summaries may still sometimes lack the precision of human-written comments.

\citeauthor{gao2022m2ts} \cite{gao2022m2ts} introduces a Transformer-based method that integrates multi-scale AST and code token features to generate accurate code summaries. Experiments on Java and Python datasets demonstrate its effectiveness. However, the approach's complexity and reliance on the quality of ASTs, along with limited dataset diversity, pose challenges to its generalizability and computational efficiency.

Unlike natural languages, source code comprehension is impacted by the grammatical connections among tokens, irrespective of their identifier name. \citeauthor{cheng2021gn}\cite{cheng2021gn} showed that the relationships between tokens that are not immediately apparent from the source code can be captured by graph representations of the source code, such as the Abstract Syntax Tree (AST). \citeauthor{cheng2021gn} provided GN-Transformer, a unique approach for end-to-end learning on a fused sequence and graph modality that is referred to as Syntax-Code-Graph (SCG). By employing a self-attention mechanism, GN-Transformer builds upon the Graph Networks (GN) foundation. An early marriage of an AST representation and a fragment of source code produced SCG. After conducting experiments on the SCG structure, an ablation research on the model design, and an analysis of the hyper-parameters, it has been determined that the fused representation offers a performance advantage. In two code summarizing datasets and across three automatic code summarization metrics (BLEU, METEOR, ROUGE-L), the suggested approaches reach state-of-the-art performance. 

They developed a new architecture called GN-Transformer by extending Graph Networks (GN). GN encoder blocks in order, then a Transformer decoder that goes from sequence to sequence. The paper introduced Syntax-Code Graph (SCG), a novel approach for the early fusion of a code snippet sequence with the related AST representation.

\begin{figure}[h]
    \centering
    \includegraphics[width=0.9\textwidth]{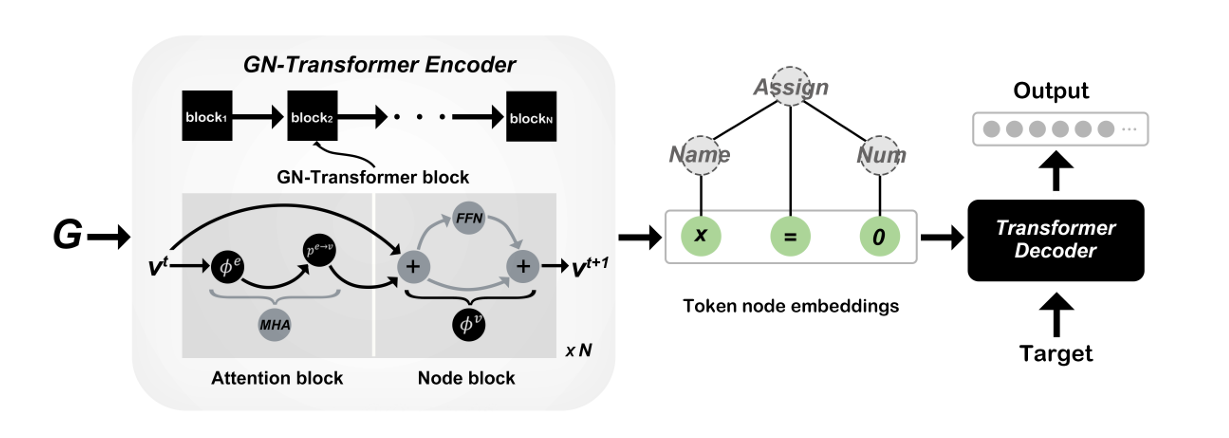}
    \caption{The encoder takes a graph $G$ as input and is composed of several GN-Transformer blocks. "+" is what is referred to as a residual connection that comes after a normalizing layer. The changed node properties are output in graph $G'$ by the encoder. Only the token nodes (green) are fed into the decoder for the code-summarization process; the AST nodes (grey) are discarded.}
    \label{fig:l1l} 
\end{figure}

\section{Large Language Models}

Incorporating Large Language Models (LLMs) into code summarization represents a significant advancement by leveraging their extensive language comprehension. Recent research has focused on utilizing these models to automatically generate concise, meaningful descriptions of code snippets, a task traditionally performed manually.

Ahmed and Devanbu \cite{ahmed2022fewshot} presented an innovative approach utilizing the GPT Codex model to perform few-shot learning for project-specific code summarization tasks. They investigated whether the few-shot capabilities of LLMs could be extended to code summarization, achieving positive results that suggest significant improvements over traditional models trained on large datasets. Their methodology involved structuring prompts with a small set of function-comment pairs before appending a target function for summarization. This set-up exploited the model's ability to generate high-quality summaries based on minimal examples tailored to the specifics of a project, highlighting the adaptability and efficiency of LLMs in handling domain-specific knowledge \cite{ahmed2022fewshot}.

The authors utilized a prompt-based approach where several function-summary pairs from the same project were presented to the model, followed by the function requiring a summary. This few-shot learning setup, without any model re-training, demonstrated that LLMs could effectively adapt to new tasks through context switching. Remarkably, the method required no weight adjustments to the model, relying instead on its inherent capacity to generate contextually appropriate responses.

The performance of the LLM, particularly in generating summaries for unseen code, was quantitatively evaluated using metrics like BLEU-4, showing improvements over state-of-the-art models. Specifically, in cross-project settings, the LLM achieved a higher BLEU score, suggesting better generalization across different coding projects. Further, same-project few-shot training enhanced the model's performance, affirming the benefit of aligning the training samples with project-specific characteristics \cite{ahmed2022fewshot}.

Haldar et al. (2024) conducted research to determine how token overlap between the source code and its descriptions affects LLM performance. To look at how performance varies between models, especially Llama 2, they split datasets into groups based on token overlap measures. Their results show that LLMs are good at recognizing syntactic patterns, but their semantic understanding is still very basic, as it is highly impacted by token overlap \cite{haldar2024performance}.

\begin{figure}[h]
    \centering
    \includegraphics[width=0.9\textwidth]{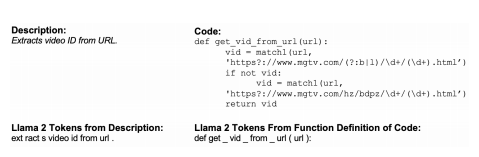}
    \caption{ The description and the first line of code are tokenized by the Llama 2 tokenizer.
This provides helpful information by displaying the tokens that are included in the description.}
    \label{fig:l1l} 
\end{figure}

In 2024, Guo et al. created GraphcodeBERT, a new way to understand semantic code that combines data flow with standard BERT architectures. GraphcodeBERT did much better at code summarization tasks than traditional models because it used execution semantics in the pre-training step. This improvement shows how important deep semantic processing is for making LLM work better on jobs that involve code \cite{guo2024graphcodebert}.

Based on multi-task learning techniques, Nijkamp et al. (2024) showed CodeT5, an encoder-decoder model that was fine-tuned for code summarization. This model was already trained on a big corpus and was fine-tuned to do the best job of summarizing code. The results from CodeT5 showed that it could make summaries that were both relevant to the context and logically consistent. This set a new standard for code summarization \cite{nijkamp2024codet5}.

%% file: chapters/methodology.tex
\chapter{Methodology}\label{chapter:methodology}
Proposing an innovative methodology for comparing LLMs in Code Summarization involves employing diverse evaluation metrics and benchmark datasets to capture different aspects of code structure. Here’s a detailed
breakdown of the methodology.
\section{Proposed Approach}
 

\begin{figure}[hbtp]
    \centering
    \includegraphics[width=1.1\textwidth]{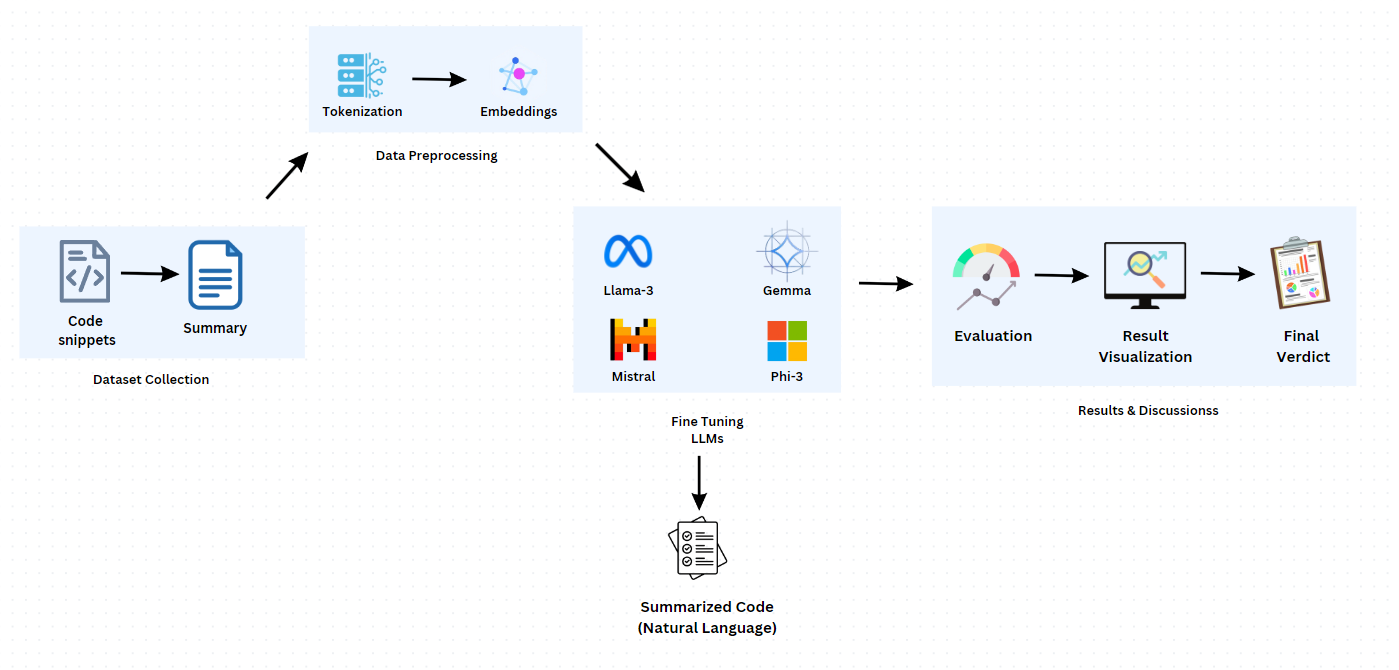}
    \caption{Proposed Methodology}
    \label{fig:proposed methodology }
\end{figure}


For investigating the answers we have employed a robust and structured methodology for our research. The overall pipeline of the methodology is illustrated in the figure~\ref{fig:proposed methodology }

\section{Dataset Collection}
For many code-NL tasks, such as code summarization, Code-XGLUE serves as a standard benchmark \cite{lu2021codexglue}. Code samples in Go, Java, JavaScript, PHP, Python, and Ruby are taken from publicly accessible, open-source GitHub projects and are accompanied by English descriptions. It is a filtered subset of CodeSearchNet \cite{husain2019codesearchnet}. The repository's opening paragraph contains a description of every code element. Examples that did not have a length of three to 256 tokens, were empty, contained special characters like "http://," had descriptions written in languages other than English, or could not be processed into an abstract syntax tree were removed from the dataset.

\begin{table}[h]
\centering
\begin{tabular}{>{\bfseries}m{3cm}|m{2cm}|m{2cm}|m{2cm}}
\hline
Languages & Train & Test & Validation \\ \hline
PHP & 241k & 14k & 13k \\ \hline
JavaScript & 58k & 3.29k & 3.89k \\ \hline
Ruby & 24.9k & 1.26k & 1.4k \\ \hline
Go & 167k & 8.12k & 7.33k \\ \hline
Java & 165k & 11k & 5.18k \\ \hline
Python & 252k & 14.9k & 13.9k \\ \hline
\end{tabular}
\caption{Experiments were conducted on the public code summarization benchmark CodeXGlue}
\label{dataset_table}
\end{table}

Table~\ref{dataset_table} presents an example taken from the code summarization dataset. We use the 251,820 training, 13,914 dev, and 14,918 testing data points from the CodeXGLUE Python examples for our research.

\section{Data Preprocessing}
The data preprocessing stage involves two critical steps: tokenization and embedding generation.
\begin{itemize}
    \item \textbf{Tokenization:} The initial stage of the data preparation pipeline involves breaking down the input text data into smaller pieces known as tokens. Depending on the tokenizer being used, these tokens may be characters, words, or subwords. Because it converts the unstructured text into a format that the model can comprehend and analyze, tokenization is crucial.
    \item \textbf{Embeddings:} After tokenization, each token is converted into a dense vector representation known as an embedding. Embeddings capture the semantic meaning of tokens in a continuous vector space, allowing the model to understand and work with the textual data more effectively. These embeddings serve as the input for subsequent layers in the model, facilitating tasks like language modeling, translation, and summarization.
\end{itemize}

\section{Model Selection}
In the new era of code summarization, various large language models (LLMs) offer distinct advantages, each suited to different needs and contexts. The models considered here are Llama-3, Phi-3, Gemma, and Mistral, each bringing unique features and capabilities to the table.

\subsection*{Llama-3\cite{llama3modelcard}}
Llama-3 has an advanced architecture that makes it very efficient in terms of processing and memory. It has a 128k tokenizer, RMSNorm, and K-V cache. It can comprehend a broad range of programming scenarios because it has been pretrained on an enormous dataset including 15 trillion tokens. Reinforcement learning from human feedback (RLHF) and supervised fine-tuning improve Llama-3 even further. It is a great option for scenarios requiring high performance and resource efficiency because its main goal is to enable faster training and efficient calculation.

\subsection*{Phi-3\cite{abdin2024phi}}
Phi-3 is designed with a similar architecture to Llama2, featuring GeGLU and an extensive 128K context window with LongRope. It is pretrained on 3.3 trillion tokens sourced from both general and specific web data, providing a robust foundation for code summarization tasks. Phi-3 undergoes supervised fine-tuning with Direct Policy Optimization (DPO), enhancing its adaptability. It is particularly suitable for deployment on handheld devices due to its efficient quantization to 4-bits, balancing performance with hardware constraints.

\subsection*{Gemma\cite{team2024gemma}}
Gemma leverages an architecture that includes Multi Query Attention, GeGLU, and RMSNorm, optimized for lightweight performance. It is pretrained on a more modest dataset of 6 trillion tokens but excels in supervised fine-tuning and reinforcement learning from feedback. This model is designed for environments where computational resources are limited, yet competitive results are required. Gemma's training on a smaller corpus makes it ideal for targeted applications where efficiency and specificity are paramount.

\subsection*{Mistral\cite{jiang2023mistral}}
Mistral features an advanced architecture with Grouped Query Attention (GQA) and Sliding Window Attention (SWA), tailored for handling long sequences effectively. It is pretrained on 3.3 trillion tokens from diverse data sources, providing a broad understanding of programming contexts. Supervised fine-tuning and Direct Policy Optimization (DPO) further refine its capabilities. Mistral is optimized for handheld device deployment, benefiting from 4-bit quantization to deliver efficient performance without compromising on the ability to handle complex tasks.
\section{Evaluation Metrics}
In our research to evaluate the performance of the LLMs we have selected the following performance metrics.
\begin{enumerate}




    \item \textbf{BLEU}:
\textit{BLEU} is a frequently used statistic in software engineering and natural language processing to evaluate generative processes, including conversation creation, Using \(n \)-gram matching, BLEU calculates the ratio of \( N \) groups of word similarity between generated comments and reference comments.

The formula is as follows: \begin{equation} \text{BLEU-N} = \text{BP} \times \exp\left(\sum_{n=1}^{N} w_n \log p_n\right), \label{BLEU} \end{equation}, in which \( P_n \) is the proportion of the candidate's subsequences that have length \(n \). The uniform weight \( 1/N \) is represented by \( \tau_n \) for short generated sequences, and the shortness penalty is represented by \( BP \). Since corpus-level BLEU-4, or \( N = 4 \), has been shown to be more correlated with human evaluations than alternative metrics, we utilize it as our assessment metric.

      \item \textbf{ROUGE-L}:
In the field of natural language processing, \textit{ROUGE-L} is frequently used to handle text summarization problems. The F-measure yields the ROUGE-L value, which is based on the Longest Common Subsequence (LCS) between two texts. ROUGE-L is calculated as follows given a generated text (X) and a reference text (Y), whose lengths are m and n, respectively:   
 \begin{equation}
P_{lcs} = \frac{LCS(X, Y)}{n}, \quad R_{lcs} = \frac{LCS(X, Y)}{m}, \quad F_{lcs} = \frac{(1 + \beta^2)P_{lcs}R_{lcs}}{R_{lcs} + \beta^2P_{lcs}},
\label{ROUGE-L}
\end{equation}

$\text{where} \ \beta = P_{lcs}/{R_{lcs}} \ \text{and} \ F_{lcs} \ \text{is calculated ROUGE-L value.}$

\end{enumerate}



    

%% file: chapters/result.tex
\chapter{Result }\label{chapter:result}

\section{Experimental Setup}
\subsection{Environment}
For model training, we utilised the free edition of Google Colab, which gave us access to NVIDIA Tesla K80 GPUs. Even though it wasn't as powerful as specialised high-performance hardware, this configuration worked well for our tests. Utilising GPU acceleration to meet the computational demands of our models, the Colab environment made it easier to train and evaluate  LLaMA-3, Phi-3, Mistral and Gemma. We conducted our trials effectively in spite of the limitations of the free version, which included limited memory capacity and session durations. This showed that employing readily available resources for complex natural language processing tasks is feasible.

\subsection{Train-Test Split}
In our comparative study, we adopted a 70-20-10 train-test-development split to ensure robust evaluation and generalization of the Large Language Models (LLMs). The training set, comprising 70\% of the dataset, is used to fit the model parameters and learn the underlying patterns in the data. This substantial portion allows the model to capture a wide variety of examples, enhancing its learning capabilities. The testing set, which constitutes 20\% of the data, is reserved for evaluating the performance of the trained models on unseen data, providing an unbiased assessment of the model's accuracy and effectiveness. Additionally, the development set, making up 10\% of the dataset, is utilized during the training process to tune hyperparameters and make model adjustments. 

\begin{table}[h!]
\centering
\begin{tabular}{c|c|c}
\hline
\textbf{Dataset} & \textbf{Percentage} & \textbf{Description} \\ \hline
Train Set        & 70\%   &   For model training          \\ 
Test Set         & 20\%   & For performance evaluation            \\ 
Dev Set          & 10\%   &  For hyperparameter tuning         \\ \hline
\end{tabular}
\caption{Train-Test-Development Split}
\label{table:traintestdev}
\end{table}

This set facilitates intermediate evaluations, ensuring the model does not overfit the training data and aiding in fine-tuning for optimal performance. By employing this 70-20-10 split, we establish a balanced and comprehensive evaluation framework that allows for effective model training, unbiased performance assessment, and fine-tuning to achieve the best results in our code summarization tasks.

\subsection{Hyper Parameters}
For the hyperparameters used in the model training process, we have used batch size of 2 is employed, meaning that the model processes two samples before updating its parameters. This small batch size helps in handling memory constraints while training. Gradient accumulation is set to 4, indicating that the gradients from four batches are accumulated before performing a backward pass. This effectively increases the batch size and stabilizes training. The warm-up steps are set to 5, gradually increasing the learning rate at the beginning of the training to prevent sudden large updates that could destabilize the model. Only one epoch is used, implying that the entire training dataset is passed through the model once. The learning rate is set to 2e-4, controlling the step size for updating model parameters during optimization. The optimizer used is AdamW 8-bit, which helps reduce memory usage while maintaining efficient and effective parameter updates.

\begin{table}[h]
\centering
\begin{tabular}{>{\bfseries}m{9cm}|m{5cm}}
\hline
Hyperparameter & Value \\ \hline
per\_device\_train\_batch\_size & 2 \\ \hline
gradient\_accumulation\_steps & 4 \\ \hline
warmup\_steps & 5 \\ \hline
num\_epochs & 1 \\ \hline
learning\_rate & 2e-4 \\ \hline
logging\_steps & 1 \\ \hline
optim & adamw\_8bit \\ \hline
weight\_decay & 0.01 \\ \hline
lr\_scheduler\_type & linear \\ \hline
seed & 3407 \\ \hline
\end{tabular}
\caption{Hyperparameters configuration for model training}
\end{table}

Overall, these hyperparameters are chosen to balance training efficiency and stability, considering memory constraints and ensuring the model learns effectively without overfitting or underfitting.

\subsection{Fine Tuning Models}
The model has to undergo through a number of crucial processes in order to be fine-tuned for maximum performance on particular tasks. Alpaca is used to format the dataset at the start of the procedure, ensuring that it is in a format that is appropriate for training. The text data is then transformed into tokens that the model can process by the tokenization stage.

Tokenization is followed by loading the model and adding a LoRA (Low-Rank Adaptation) adaptor to improve the model's training efficiency and flexibility. The prepared dataset is then used to train the model using the SFTTrainer (Supervised Fine-Tuning Trainer). In order to increase performance on the goal task, this stage entails modifying the model parameters depending on the training data.

The Adam optimizer is used to maximize the training process. This optimizer is well-known for its effectiveness when working with sparse gradients, and it works especially well when fine-tuning big models. In order to assess the model's performance and make sure it satisfies the required relevance and accuracy criteria, the inference phase is finally carried out using test data.


\section{Performance Result}

\subsection{Performance Evaluation on Python Dataset}
The performance evaluation of four large language models (LLMs) utilizing BLEU and ROUGE-L scores on the Python dataset is shown in table \ref{tab:python}. Gemma-7b, Phi-3-medium, Llama-3-8b, and Mistral-7b are the models that have been assessed. The highest BLEU scores of 7.38 are obtained by Phi-3-medium and Mistral-7b, suggesting that their generated summaries have the maximum degree of n-gram overlap with the reference summaries. These two models also obtain the maximum ROUGE-L score of 19.35, indicating that in addition to being correct, their summaries are also fluent and pertinent to the context. Conversely, Gemma-7b and Llama-3-8b exhibit a respectable performance, but somewhat lagging below Phi-3-medium and Mistral-7b, with BLEU scores of 7.23 and ROUGE-L scores of 18.95, respectively.

\begin{table}[h]
\centering
\begin{tabular}{c|c|c}
\hline
\rowcolor{blue!20}
\textbf{Models} & \textbf{BLEU} & \textbf{ROUGE-L} \\
\hline
Gemma-7b & 7.23 & 18.95 \\
Phi-3-medium & 7.38 & 19.35 \\
Llama-3-8b & 7.23 & 18.95 \\
Mistral-7b & 7.38 & 19.35 \\
\hline
\end{tabular}
\caption{Evaluation Metrics for Python Dataset}
\label{tab:python}
\end{table}

\subsection{Performance Evaluation on Java Dataset}
Besides the table \ref{tab:java} shows the performance evaluation of the same four models on the Java dataset, again using BLEU and ROUGE-L scores. For this dataset, the results differ somewhat. Llama-3-8b achieves the highest BLEU score of 6.10, indicating it produces the most n-gram similar summaries to the reference text. However, Mistral-7b outperforms in terms of ROUGE-L score with a value of 22.37, suggesting its summaries are more contextually appropriate and fluent. Gemma-7b and Phi-3-medium, while showing lower BLEU scores of 6.02 and 5.17 respectively, still perform competitively in terms of ROUGE-L scores, with Phi-3-medium achieving 21.82 and Gemma-7b at 21.11. These results illustrate that Mistral-7b consistently generates relevant and fluent summaries, while Llama-3-8b excels in n-gram similarity for the Java dataset.

\begin{table}[h]
\centering
\begin{tabular}{c|c|c}
\hline
\rowcolor{blue!20}
\textbf{Models} & \textbf{BLEU} & \textbf{ROUGE-L} \\
\hline
Gemma-7b & 6.02 & 21.11 \\
Phi-3-medium & 5.17 & 21.82 \\
Llama-3-8b & 6.10 & 20.12 \\
Mistral-7b & 5.71 & 22.37 \\
\hline
\end{tabular}
\caption{Evaluation Metrics for Java Dataset}
\label{tab:java}
\end{table}

\subsection{Performance Evaluation on Go Dataset}
The table \ref{tab:go} evaluates the performance of four large language models (LLMs)—Gemma-7b, Phi-3-medium, Llama-3-8b, and Mistral-7b—on the Go dataset using BLEU and ROUGE-L scores. Gemma-7b achieves a BLEU score of 1.01 and a ROUGE-L score of 7.14, indicating relatively lower performance compared to the other models. Both Phi-3-medium and Llama-3-8b achieve the highest BLEU score of 1.37 and the highest ROUGE-L score of 7.69, suggesting that they produce summaries with better n-gram overlap and contextual relevance than Gemma-7b and Mistral-7b. Mistral-7b scores slightly lower than Phi-3-medium and Llama-3-8b, with a BLEU score of 1.35 and a ROUGE-L score of 7.48, still performing well but not the best in this dataset.

\begin{table}[h]
\centering
\begin{tabular}{c|c|c}
\hline
\rowcolor{blue!20}
\textbf{Models} & \textbf{BLEU} & \textbf{ROUGE-L} \\
\hline
Gemma-7b & 1.01 & 7.14 \\
Phi-3-medium & 1.37 & 7.69 \\
Llama-3-8b & 1.37 & 7.69 \\
Mistral-7b & 1.35 & 7.48 \\
\hline
\end{tabular}
\caption{Evaluation Metrics for Go Dataset}
\label{tab:go}
\end{table}

\subsection{Performance Evaluation on JavaScript Dataset}
The table \ref{tab:javascript} evaluates the same models on the JavaScript dataset using BLEU and ROUGE-L scores. Gemma-7b scores a BLEU of 5.08 and a ROUGE-L of 17.14, which are comparatively lower than the other models in this dataset. Phi-3-medium achieves a BLEU score of 7.49 and a ROUGE-L score of 30.09, demonstrating strong performance in generating both accurate and contextually relevant summaries. Llama-3-8b scores the highest BLEU at 7.76, indicating it produces summaries with the greatest n-gram overlap with the reference text, and a ROUGE-L of 24.00. However, Mistral-7b outperforms all models with a BLEU score of 13.79 and a ROUGE-L score of 36.84, indicating it generates the most accurate and contextually relevant summaries for JavaScript.

\begin{table}[h]
\centering
\begin{tabular}{c|c|c}
\hline
\rowcolor{blue!20}
\textbf{Models} & \textbf{BLEU} & \textbf{ROUGE-L} \\
\hline
Gemma-7b & 5.08 & 17.14 \\
Phi-3-medium & 7.49 & 30.09 \\
Llama-3-8b & 7.76 & 24.00 \\
Mistral-7b & 13.79 & 36.84 \\
\hline
\end{tabular}
\caption{Evaluation Metrics for JavaScript Dataset}
\label{tab:javascript}
\end{table}

\subsection{Performance Evaluation on PHP Dataset}
Lastly the table \ref{tab:php} evaluates the performance of four large language models (LLMs) on the PHP dataset using BLEU and ROUGE-L scores. The models assessed are Gemma-7b, Phi-3-medium, Llama-3-8b, and Mistral-7b. Gemma-7b achieves a BLEU score of 3.87 and a ROUGE-L score of 12.66, indicating moderate performance. Phi-3-medium slightly outperforms Gemma-7b with a BLEU score of 4.03 and a ROUGE-L score of 12.82, suggesting marginally better n-gram overlap and contextual relevance. Llama-3-8b performs significantly better with a BLEU score of 6.24 and a ROUGE-L score of 13.82, demonstrating its strength in generating accurate and relevant summaries. Mistral-7b stands out with the highest BLEU score of 8.79 and a ROUGE-L score of 12.99, indicating its superior ability to produce precise and contextually appropriate summaries for the PHP dataset.

\begin{table}[h]
\centering
\begin{tabular}{c|c|c}
\hline
\rowcolor{blue!20}
\textbf{Models} & \textbf{BLEU} & \textbf{ROUGE-L} \\
\hline
Gemma-7b & 3.87 & 12.66 \\
Phi-3-medium & 4.03 & 12.82 \\
Llama-3-8b & 6.24 & 13.82 \\
Mistral-7b & 8.79 & 12.99 \\
\hline
\end{tabular}
\caption{Evaluation Metrics for PHP Dataset}
\label{tab:php}
\end{table}

\subsection{Performance Evaluation on Ruby Dataset}
The second table assesses the same models on the Ruby dataset using BLEU and ROUGE-L scores. Gemma-7b achieves a BLEU score of 3.53 and a ROUGE-L score of 8.33, reflecting moderate performance. Phi-3-medium emerges as the best performer with a BLEU score of 5.03 and a notably high ROUGE-L score of 21.35, indicating it generates summaries with the highest contextual relevance and fluency. Llama-3-8b also performs well with a BLEU score of 4.38 and a ROUGE-L score of 8.39, showing its capability to generate accurate summaries. In contrast, Mistral-7b has the lowest BLEU score of 2.52 and a ROUGE-L score of 7.74, suggesting it struggles more with the Ruby dataset compared to the other models.

\begin{table}[h]
\centering
\begin{tabular}{c|c|c}
\hline
\rowcolor{blue!20}
\textbf{Models} & \textbf{BLEU} & \textbf{ROUGE-L} \\
\hline
Gemma-7b & 3.53 & 8.33 \\
Phi-3-medium & 5.03 & 21.35 \\
Llama-3-8b & 4.38 & 8.39 \\
Mistral-7b & 2.52 & 7.74 \\
\hline
\end{tabular}
\caption{Evaluation Metrics for Ruby Dataset}
\label{tab:ruby}
\end{table}


\section{Result Visualization}

The following analysis discusses the performance of four large language models (LLMs)—Gemma, Llama-3, Mistral, and Phi-3—based on BLEU and ROUGE-L scores across different programming language datasets: JavaScript, Java, Go, Ruby, Python, and PHP.

\subsection*{BLEU Scores of Models Across Different Datasets}

\begin{figure}[htbp]
    \centering
    \includegraphics[width=0.7\textwidth]{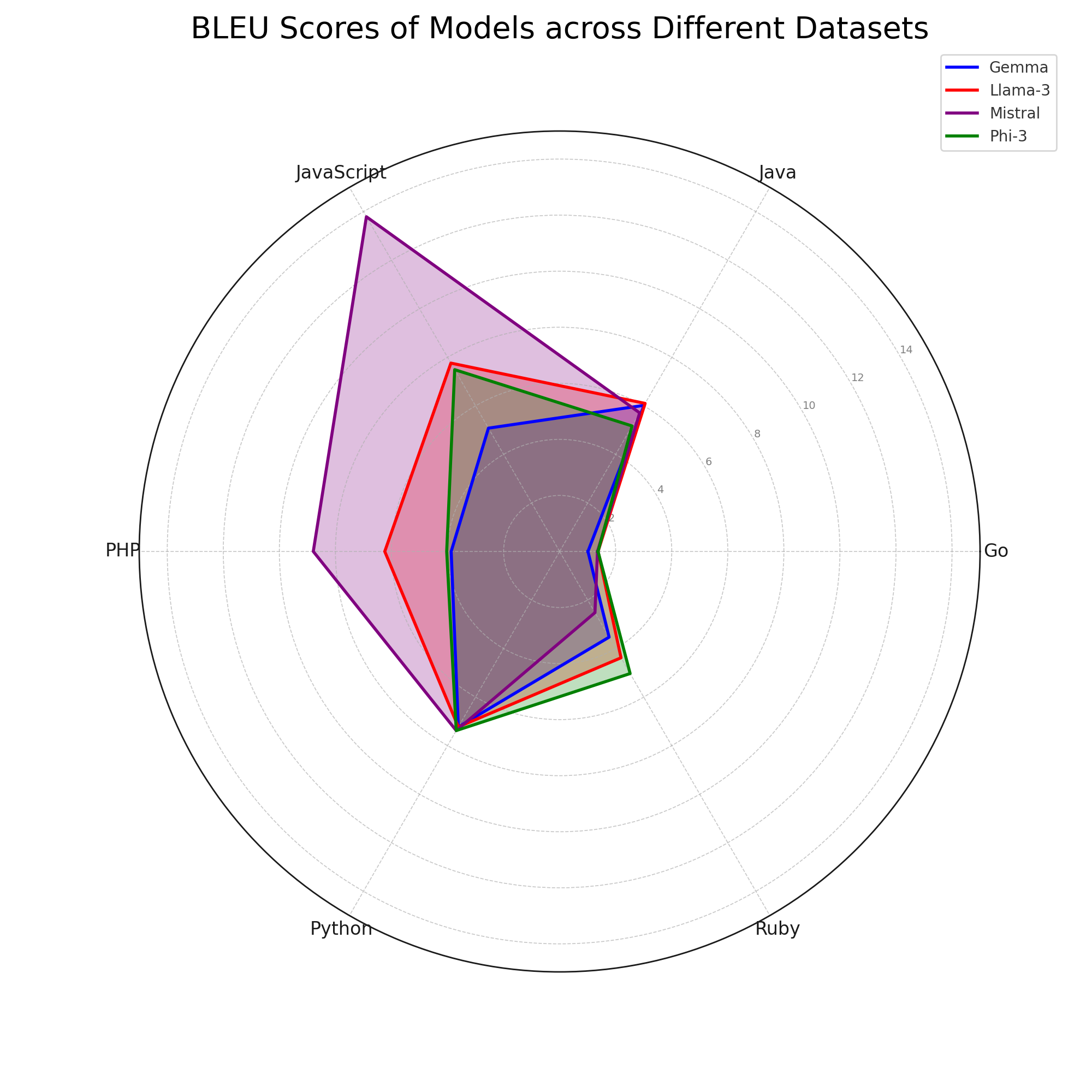}
    \caption{Visualization for LLMs Performance in BLEU scores}
    \label{fig:bleu_scores} 
\end{figure}

The radar chart \ref{fig:bleu_scores} displays the BLEU scores of the models across various datasets. Key observations include:
\begin{itemize}
    \item \textbf{Mistral} (in red) consistently achieves high BLEU scores, particularly excelling in the JavaScript and PHP datasets. This indicates that Mistral generates text with the greatest n-gram overlap with reference texts in these languages.
    \item \textbf{Llama-3} (in blue) also performs well, with notable scores in the JavaScript and Java datasets.
    \item \textbf{Phi-3} (in green) demonstrates strong performance in the Ruby dataset, outperforming the other models in this specific language.
    \item \textbf{Gemma} (in purple) tends to have lower BLEU scores across most datasets compared to the other models, indicating relatively less n-gram similarity with the reference texts.
\end{itemize}

\subsection*{ROUGE-L Scores of Models Across Different Datasets}

\begin{figure}[htbp]
    \centering
    \includegraphics[width=0.7\textwidth]{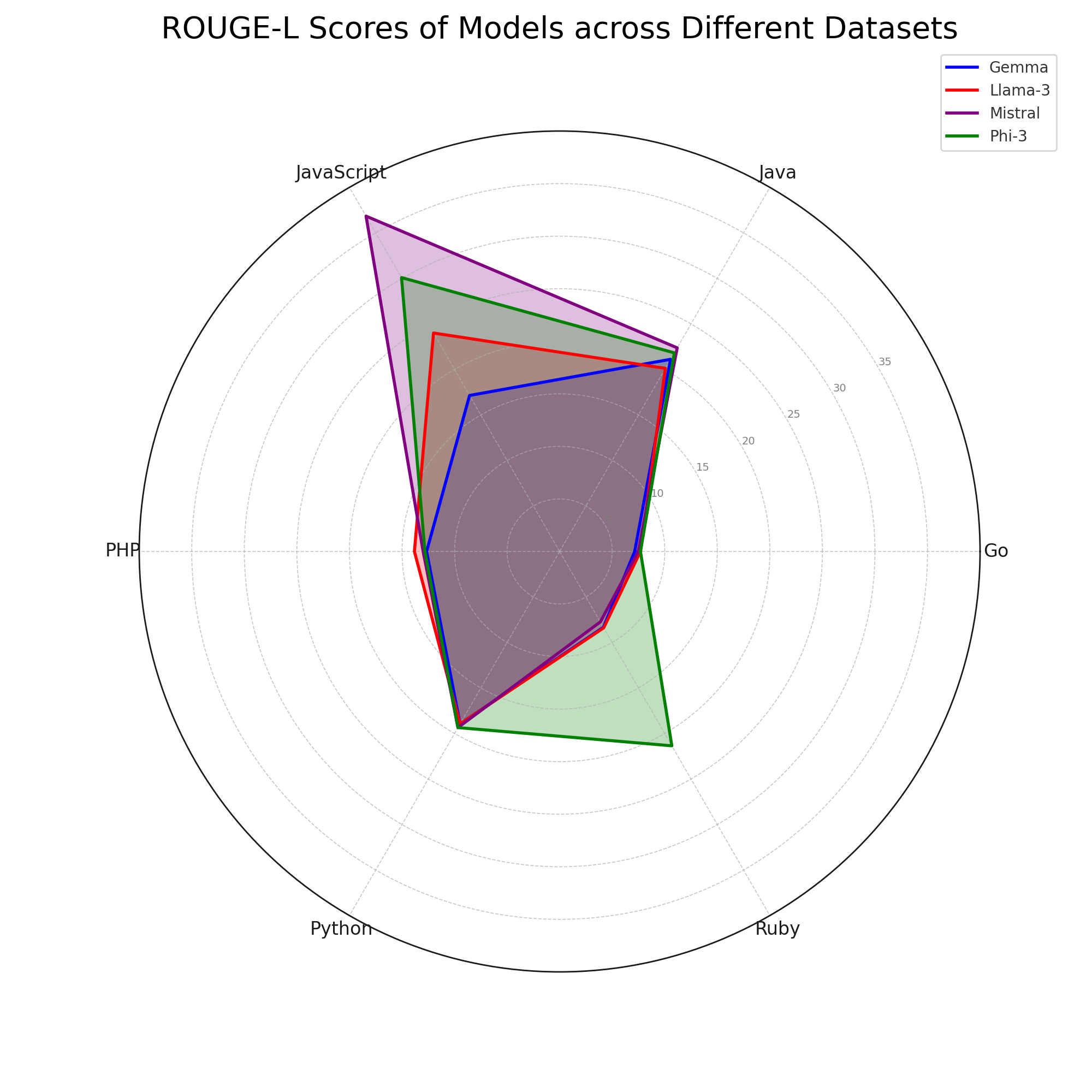}
    \caption{Visualization for LLMs Performance in ROUGE-L scores}
    \label{fig:rouge_scores} 
\end{figure}

The ROUGE-L scores for the same models over the same datasets are displayed in the chart \ref{fig:rouge_scores}, which measures the longest common subsequence (LCS) between generated text and reference text. Important findings consist of:
\begin{itemize}
    \item \textbf{Mistral} (in red) again stands out with high ROUGE-L scores, particularly in the JavaScript and PHP datasets, showcasing its ability to produce contextually relevant and fluent summaries.
    \item \textbf{Phi-3} (in green) excels in the Ruby dataset, highlighting its strong performance in generating summaries with high contextual relevance and fluency for Ruby code.
    \item \textbf{Llama-3} (in blue) performs well in the JavaScript and Java datasets, indicating its capability in these languages.
    \item \textbf{Gemma} (in purple) shows lower ROUGE-L scores across most datasets, suggesting it is less effective in generating contextually coherent summaries compared to the other models.
\end{itemize}

\section{Final Verdict}

In our comparative analysis of large language models for code summarization, Mistral emerges as the best overall performer, demonstrating high scores across most datasets, with exceptional performance in JavaScript and PHP. Phi-3 also performs consistently well, particularly excelling in Ruby and Go. While Llama-3 and Gemma exhibit decent performances in specific datasets, they generally lag behind the leading models, Mistral and Phi-3. These findings highlight the strengths of Mistral and Phi-3 in delivering high-quality code summaries across a variety of programming languages.

\section{Limitation}
Each model may require individualized training strategies to accommodate differences in architecture and learning dynamics, despite being subjected to the same initial conditions. The availability and quality of training data are critical yet challenging, as high-quality, domain-specific datasets for code summarization are scarce. This scarcity can hinder the models' ability to generalize effectively to diverse real-world coding practices. Additionally, ensuring that each model's output is not only accurate but also contextually relevant to developers is a complex task that involves balancing technical accuracy with practical usability. These challenges highlight the need for a nuanced approach to training and evaluating each model within your research framework.

Another critical limitation is the potential for model bias. Since LLMs are trained on available data, any inherent biases in this data can be amplified during training and fine-tuning. This could lead to models that perform well on certain types of code or coding styles but poorly on others, potentially affecting the fairness and inclusivity of the tool.

Code practices and programming languages evolve, so models trained on current datasets may quickly become outdated. Continuously updating the models to adapt to new programming paradigms or languages is a significant challenge, requiring ongoing collection of new data and retraining.

%% file: chapters/discussion.tex
\chapter{Future Work and Conclusion}\label{chapter:discussion}

\section{Future Work}

The future direction for the comparative analysis of large language models (LLMs) in code summarization should encompass several key areas. First, expanding the evaluation to include a broader range of LLMs will provide a more comprehensive performance comparison, allowing for a better understanding of the strengths and weaknesses of different models. This can help in identifying the most suitable models for specific code summarization tasks. Second, enhancing the model's understanding of code semantics, beyond mere syntax, is crucial. Improving the semantic comprehension will lead to higher quality summaries that accurately capture the intent and functionality of the code, rather than just its structure. Lastly, focusing on reducing latency and computational requirements is essential to make the models more efficient and practical for real-world applications. This can be achieved through optimization techniques and more efficient model architectures, ensuring that the benefits of advanced LLMs are accessible without prohibitive computational costs.  

\newpage
\section{Conclusion}

In conclusion, our evaluation of various large language models (LLMs) for code summarization reveals that Mistral-7b and Phi-3-medium are the top performers across multiple datasets. Mistral-7b consistently demonstrates superior capabilities, particularly excelling in JavaScript and PHP, while Phi-3-medium shows strong performance, especially in Ruby and Go. Llama-3-8b and Gemma-7b also produce decent results in specific datasets but generally fall behind the leading models. These findings highlight the importance of selecting the appropriate model based on the specific programming language and context. Future work should focus on further enhancing the understanding of code semantics and optimizing model efficiency to improve performance and applicability in real-world scenarios.